\documentclass[pra,twocolumn,showpacs,floatfix,aps]{revtex4}
\usepackage{graphicx}

\begin{document}

\title{Marching toward the eigenvalues: The Canonical Function Method and the Schr\"odinger equation}
\author{C. Tannous and J. Langlois}
\affiliation{Laboratoire de Magn\'etisme de Bretagne, CNRS-FRE 3117,
Universit\'e de Bretagne Occidentale, BP: 809 Brest CEDEX, 29285 FRANCE}

\date{February, 2010}

\begin{abstract}
The Canonical Function Method (CFM) is a powerful accurate and fast method that solves
the Schr\"{o}dinger equation for the eigenvalues directly without having to evaluate the
eigenfunctions. Its versatility allows to solve several types of problems and in 
this work it is applied to the solution of several 1D potential problems,
the 3D Hydrogen atom and the Morse potential.
\end{abstract}

\pacs{03.65.-w; 31.15.Gy; 33.20.Tp; 03.65.Nk; 02.60.Cb}

\maketitle

\section{Introduction}

The Canonical Function Method (CFM) is a powerful means for solving the Schr\"{o}dinger
equation and getting the eigenvalue spectrum directly in a fast and precise manner without
computing the eigenfunctions.

The CFM turns the two-point boundary value (TPBV) Schr\"{o}dinger problem into an initial 
value problem and allows full and accurate determination of the spectrum. 
This is done by expressing the solution as a sum of two linearly independent functions (the
Canonical Functions) with specific values at some arbitrary point belonging to the interval
defined by the two boundaries. 
The integration proceeds simultaneously from this point
toward the left and right boundaries evaluating at each step a corresponding ratio. It stops when
the difference between the left and right ratios is below a given desired precision. 

This work is relevant to students who have completed an undergraduate Quantum Mechanics course 
of the Merzbacher~\cite{Merzbacher} level  
or graduate students whose level corresponds to Landau and Lifshitz course~\cite{Landau} and are interested in the eigenvalue problem of Quantum Mechanics. 

The CFM can handle a large variety of Quantum problem problems~\cite{Tannous} besides the eigenvalue
problem making it an extremely versatile, fast  and highly accurate.
The evaluation of the Schr\"{o}dinger operator spectrum is done without performing
diagonalization, bypassing the evaluation of the eigenfunctions. This
allows to preserve a high degree of numerical precision that is required in solving sensitive eigenvalue problems. 

It also solves the Radial Schr\"{o}dinger
equation over the infinite interval $[0, \infty[$ , where singularities in the potential at both
boundaries are encountered.

The CFM method is superior to many standard techniques that have been used to solve the
Schr\"odinger equation such as Numerov~\cite{Numerov} or relaxation methods that are
particularly tailored for solving TPBV problems (see the Physics Reports Review~\cite{Tannous}).

It is worthwhile to point out that, numerically, the precision gained with the bypass of
 intermediate diagonalization operations
is reminiscent of the Golub-Reinsch algorithm (see for instance ref.~\cite{Recipes})
used for the singular value decomposition of arbitrary rectangular matrices.

This article is organised as follows: The next section is a description of the
CFM in 1D with the appropriate boundary conditions (BC). Several 1D problems
are treated: The Infinite depth potential well, the finite depth potential well, 
the Harmonic Oscillator problem, the Kronig-Penney potential and the 
double-well (symmetric and asymmetric) potentials. In section III the 
CFM is applied to the 3D Schr\"odinger equation specializing to the
Radial Schr\"odinger equation problems and in particular to
the Hydrogen atom and the Morse potential. Finally section IV bears our conclusions.

In the Appendix we provide information on the different systems of units used
in Atomic physics and quantum mechanics. 

\section{One-dimensional potential problems}
The CFM approach is based on the direct calculation of the eigenvalues of the 
Schr\"{o}dinger equation defined over an interval $[x_1,x_2]$ with a set of
BC defining the problem:
\begin{equation}
[-\frac{\hbar^2}{2m_e}\frac{d^2}{d x^2}+V(x)] \psi(x)=E \psi(x), \hspace{2mm} x_1 \le x \le x_2 
\end{equation} 

$m_e$ is the electron mass.

Starting from a point $x_0 \in [x_1,x_2]$ we express the solution as a superposition of two linearly 
independent functions  $\alpha(E;x), \beta(E;x)$, (the Canonical Functions) depending on the energy $E$ and the
abscissa $x$   such that:
\begin{equation}
y(x)= y(x_0) \alpha(E;x)  + y'(x_0)\beta(E;x)
\end{equation} 

The CFM is based on the extraction of the eigenvalues from the zeroes of the
eigenvalue function $F(E)$ defined from the saturation of the left ($x \rightarrow x_1$)
and right ($x \rightarrow x_2$) functions $l_{-}(E)$ and $l_{+}(E)$ given
by the ratios of the canonical functions $\alpha(E;x)$ and $\beta(E;x)$ or their
derivatives. 

In the general case when either $y(x_1) \ne 0 $  or $y(x_2) \ne 0$ we write:
\begin{eqnarray}
y(x)&= y(x_0) \alpha(E;x)  + y'(x_0)\beta(E;x)  \nonumber \\
y'(x)&= y(x_0) \alpha'(E;x)  + y'(x_0)\beta'(E;x)
\label{cano}
\end{eqnarray}

The canonical functions satisfy the following conditions at the starting point $x_0$:
\begin{eqnarray}
\alpha(E;x_0)=1,\alpha'(E;x_0)=0, \nonumber \\ 
\beta(E;x_0)=0,\beta'(E;x_0)=1 
\end{eqnarray}

Let us rewrite the system \ref{cano} at the two boundaries $x=x_1$ and $x=x_2$:
\begin{eqnarray}
y(x_1)&= & y(x_0) \alpha(E;x_1)  + y'(x_0)\beta(E;x_1);  \nonumber \\
y'(x_1)&=& y(x_0) \alpha'(E;x_1)  + y'(x_0)\beta'(E;x_1);  \nonumber \\
y(x_2)&=& y(x_0) \alpha(E;x_2)  + y'(x_0)\beta(E;x_2);      \nonumber \\
y'(x_2)&=& y(x_0) \alpha'(E;x_2)  + y'(x_0)\beta'(E;x_2)
\label{zero}
\end{eqnarray}

Extracting from above the left and right ratios defining the functions $l_{-}(E)$ and $l_{+}(E)$:
\begin{eqnarray}
l_{-}(E)=\left[ \frac{y'(x_0)}{y(x_0)}\right]_{-} =\frac{\alpha(E;x_1)y'(x_1)-\alpha'(E;x_1)y(x_1)}{\beta'(E;x_1)y(x_1)-\beta(E;x_1)y'(x_1)}; \nonumber \\
l_{+}(E)=\left[  \frac{y'(x_0)}{y(x_0)}\right]_{+} =
\frac{\alpha(E;x_2)y'(x_2)-\alpha'(E;x_2)y(x_2)}{\beta'(E;x_2)y(x_2)-\beta(E;x_2)y'(x_2)} \nonumber \\
\end{eqnarray}

In order to tackle any problem with the CFM, a number of constraints 
should be explained and underlined in order to illustrate the methodology of getting properly
the eigenvalue spectrum:

\begin{itemize}

\item  Sensitivity, stability and accuracy: \\
The spectrum depends on the zeroes of $F(E)=l_{+}(E)-l_{-}(E)$.
This subtraction might lead in some cases to inaccuracies because the entire spectrum
depends on the zeroes of $F(E)$. However, it holds the key of the stability of the CFM
since two independent solution sets are generated at the point $x_0$ , with progress
inwards to the left point $x_1$ and outwards toward the right point $x_2$. 
Since both sets contain, in general, linear combinations of the regular and 
the irregular solutions, by suitably combining them, the irregular solution
is eliminated.

\item $x_{0}$ issue and the number of eigenvalues: \\
The number of eigenvalues depend strongly on $x_{0}$.
Thus, it should be chosen such that a $\tan(E)$-like diagram for the energy
function $F(E)$ is obtained. In the case we have a potential displaying
a single minimum, $x_{0}$ should be close to the potential minimum. 

\item  Behaviour of the canonical functions: \\
The method being  sensitive to convergence of the marching toward
the left-right boundaries $x_1, x_2$, one ought to look for similar behaviour 
in the canonical functions $\alpha(E;x)$ and $\beta(E;x)$ along with the limiting process 
$x\rightarrow x_1$ and $x\rightarrow x_2$ since it controls the ratio saturation.

\item Overall aspect of the eigenvalue function: \\
The eigenvalue function $F(E)=l_{+}(E)-l_{-}(E)$ should have a regular structure
of the $\tan(E)$ type, that is almost periodic versus $\ln(E)$. 

\end{itemize}

There are several types of BC from the eigenvalue function defined as the difference between left and right ratio
functions: 
\begin{equation}
F(E)=l_{+}(E)-l_{-}(E)= \left[ \frac{y'(x_0)}{y(x_0)}\right]_{+} - \left[\frac{y'(x_0)}{y(x_0)}\right]_{-}
\end{equation}

We consider, for illustration,  the following four types of BC:

\begin{enumerate}
\item Null wavefunctions BC:  \\
The conditions $y(x_1)=y(x_2)=0$ yield:

\begin{eqnarray}
l_{-}(E) = \lim_{x \rightarrow x_1} -\frac{\alpha(E;x)}{\beta(E;x)}; \nonumber \\
l_{+}(E) = \lim_{x \rightarrow x_2} -\frac{\alpha(E;x)}{\beta(E;x)}  
\end{eqnarray}

\item Null wavefunction and its derivative BC:  \\
The conditions $y(x_1)=y'(x_2)=0$ yield:

\begin{eqnarray}
l_{-}(E)= \lim_{x \rightarrow x_1} -\frac{\alpha(E;x)}{\beta(E;x)}; \nonumber \\
l_{+}(E)= \lim_{x \rightarrow x_2} -\frac{\alpha'(E;x)}{\beta'(E;x)}  
\end{eqnarray}

\item Null derivative and the wavefunction BC:  \\
The conditions $y'(x_1)=y(x_2)=0$ yield:

\begin{eqnarray}
l_{-}(E)= \lim_{x \rightarrow x_1} -\frac{\alpha'(E;x)}{\beta'(E;x)};  \nonumber \\
l_{+}(E)= \lim_{x \rightarrow x_2} -\frac{\alpha(E;x)}{\beta(E;x)}   
\end{eqnarray}

\item Null derivatives BC:  \\
The conditions $y'(x_1)=y'(x_2)=0$ yield:

\begin{eqnarray}
l_{-}(E)= \lim_{x \rightarrow x_1} -\frac{\alpha'(E;x)}{\beta'(E;x)};  \nonumber \\
l_{+}(E)= \lim_{x \rightarrow x_2} -\frac{\alpha'(E;x)}{\beta'(E;x)}   
\end{eqnarray}

\end{enumerate}

\begin{figure}[!htbp]
\begin{center}
\scalebox{0.3}{\includegraphics[angle=-90]{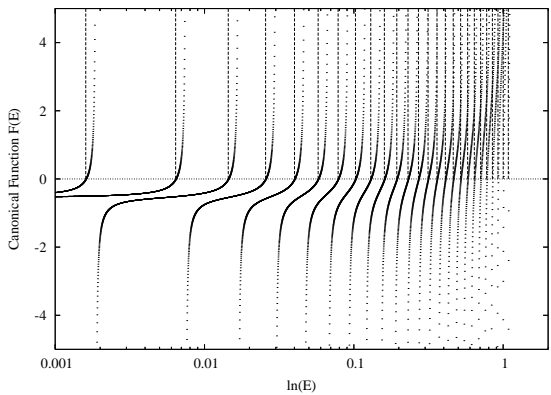}}
\end{center}
\caption{\label{tan} Eigenvalue function versus energy displaying the first 25 eigenvalues of 
the Infinite square well. The vertical lines indicate
the eigenvalue position. The eigenvalue function has an approximate
$\tan(E)$ shape versus the energy $E$.}
\end{figure}

It is remarkable that the eigenvalue function  $F(E)=l_{+}(E)-l_{-}(E)$ behaves in a very
peculiar way close to the trigonometric $\tan(E)$ shape as displayed in Fig. ~\ref{tan}. This
will be explained in the next section.

\subsection{The Infinitely deep square well}

Let us apply the CFM to the infinite square well potential of width $a$ defined by: \\
$V(x)=0, \hspace{2mm}  0 < x < a , V(0)= \infty,  V(a)= \infty$, meaning $x_1=0, x_2=a$.

The Schr\"odinger equation writes:
\begin{equation}
-\frac{\hbar^2}{2m_e}\frac{d^2 \psi(x)}{d x^2}=E \psi(x), \hspace{2mm} 0 < x < a
\end{equation} 
with BC: $ \psi(0)= 0,  \psi(a)= 0$.
The exact eigenfunctions are hence given by $\psi_n(x) = \sqrt{\frac{2}{a}}  \sin(\frac{n \pi x}{a}), n=1,2...$
yielding the exact eigenvalues as: $E_n=\frac{\hbar^2}{2m_e}{(\frac{n\pi}{a})}^2$
with $m_e$ the electron mass. \\
In order to apply, the CFM method, we first note that the solutions are odd or even
over the interval $[0,a]$. Working on the half-interval $[0,\frac{a}{2}]$ we can start
from any point $x_0$ and apply the general methodology albeit with a 
modification regarding the matching conditions at the middle interval point.

In the odd-mode case (null wavefunctions at both boundaries $x_1=0,x_2=\frac{a}{2}$):

\begin{eqnarray}
l_{-}(E)= \lim_{x \rightarrow 0} -\frac{\alpha(E;x)}{\beta(E;x)}; \nonumber \\
l_{+}(E)= \lim_{x \rightarrow \frac{a}{2}} -\frac{\alpha(E;x)}{\beta(E;x)}; \nonumber \\
F_o(E)=l_{+}(E)-l_{-}(E)
\end{eqnarray}

whereas in the even-mode case we have (null wavefunction at left boundary $x_1=0$, null 
derivative at right boundary $x_2=\frac{a}{2}$): 
\begin{eqnarray}
l_{-}(E)= \lim_{x \rightarrow 0} -\frac{\alpha(E;x)}{\beta(E;x)}; \nonumber \\
l_{+}(E)= \lim_{x \rightarrow \frac{a}{2}} -\frac{\alpha'(E;x)}{\beta'(E;x)}; \nonumber \\
F_e(E)=l_{+}(E)-l_{-}(E) 
\end{eqnarray}

Solving successively $F_o(E)$ and $F_e(E)$ for the odd modes and the even modes, we get Table~\ref{well}.

\begin{table}[!htbp]
\begin{center}
\begin{tabular}{ c c c }
\hline  Index &  CFM & Exact  \\ \hline
           1 &  1.6006952(-3) &  1.6000001(-3)    \\
           3 &  1.4406255(-2) &  1.4400001(-2)    \\
           5 &  4.0017359(-2) &  4.0000003(-2)    \\
           7 &  7.8434058(-2) &  7.8400001(-2)    \\
           9 &  0.1296562     &  0.1296000        \\
          11 &  0.1936839     &  0.1936000        \\
          13 &  0.2705172     &  0.2704000        \\
          15 &  0.3601561     &  0.3600000        \\
          17 &  0.4626004     &  0.4624000        \\
          19 &  0.5778504     &  0.5776000        \\
          21 &  0.7059059     &  0.7056000        \\
          23 &  0.8467670     &  0.8464000        \\
          25 &   1.000433     &   1.000000        \\
\hline
           2 &  6.4027691(-3) &  6.4000003(-3)    \\
           4 &  2.5611134(-2) &  2.5600001(-2)    \\
           6 &  5.7624962(-2) &  5.7600003(-2)    \\
           8 &  0.1024444     &  0.1024000        \\
          10 &  0.1600693     &  0.1600000        \\
          12 &  0.2304998     &  0.2304000        \\
          14 &  0.3137361     &  0.3136000        \\
          16 &  0.4097775     &  0.4096000        \\
          18 &  0.5186247     &  0.5184000        \\
          20 &  0.6402774     &  0.6400000        \\
          22 &  0.7747357     &  0.7744000        \\
          24 &  0.9219995     &  0.9216000        \\
\hline
\end{tabular}
\caption{\label{well} First twenty-five odd and even quantum levels of the infinite square well potential given by the CFM along with exact
results. The well width $a$ is chosen in a way such that the eigenvalue is 1 when the level index is 25. The numbers in parenthesis represent a power of 10. All eigenvalues are in 
Atomic units (see Appendix).} 
\end{center}
\end{table}

\subsection{The Finite depth square well}

The finite depth square well potential of width $a$ is defined by: 
$V(x)=0, \hspace{2mm} \mbox{for} \hspace{2mm}  0 < x < a ; \hspace{2mm} V(x)= V_0;  \hspace{2mm} \mbox{for} \hspace{2mm}  x \ge a \hspace{2mm} \mbox{or}  \hspace{2mm} x \le 0 $.

As in the Infinite depth case, the potential is symmetric with respect to the well center $\frac{a}{2}$, implying that we have odd and even
modes. Therefore we take $x_1=-\infty, x_2=\frac{a}{2}$ which means that we march to the
left through the potential step until we observe the nulling of the wavefunction,
whereas the marching to the right results at half the potential well width $a$
in odd or even modes. More explicitly: \\ 

In the odd-mode case (null wavefunctions at both boundaries $x_1=-\infty,x_2=\frac{a}{2}$):

\begin{eqnarray}
l_{-}(E)= \lim_{x \rightarrow -\infty} -\frac{\alpha(E;x)}{\beta(E;x)}; \nonumber \\
l_{+}(E)= \lim_{x \rightarrow \frac{a}{2}} -\frac{\alpha(E;x)}{\beta(E;x)}; \nonumber \\
F_o(E)=l_{+}(E)-l_{-}(E)
\end{eqnarray}

whereas in the even-mode case we have (null wavefunction at left boundary $x_1=-\infty$, null 
derivative at right boundary $x_2=\frac{a}{2}$): 
\begin{eqnarray}
l_{-}(E)= \lim_{x \rightarrow -\infty} -\frac{\alpha(E;x)}{\beta(E;x)}; \nonumber \\
l_{+}(E)= \lim_{x \rightarrow \frac{a}{2}} -\frac{\alpha'(E;x)}{\beta'(E;x)}; \nonumber \\
F_e(E)=l_{+}(E)-l_{-}(E) 
\end{eqnarray}

Solving successively $F_o(E)$ and $F_e(E)$ for the odd modes and the even modes, we get the following table~\ref{fwell}.

\begin{table}[!htbp]
\begin{center}
\begin{tabular}{ c c c }
\hline  Index &  CFM & Exact  \\ \hline
           1 &  1.6482281(-3) &  1.6475233(-3)    \\
           3 &  1.4832322(-2) &  1.4826014(-2)    \\
           5 &  4.1191306(-2) &  4.1173782(-2)    \\
           7 &  8.0705732(-2) &  8.0671579(-2)    \\
           9 &  0.1333447     &  0.1332882        \\
          11 &  0.1990615     &  0.1989772        \\
          13 &  0.2777881     &  0.2776708        \\
          15 &  0.3694246     &  0.3692691        \\
          17 &  0.4738183     &  0.4736193        \\
          19 &  0.5907167     &  0.5904695        \\
          21 &  0.7196453     &  0.7193471        \\
          23 &  0.8594448     &  0.8590948        \\
\hline
           2 &  6.5926472(-3) &  6.5898113(-3)    \\
           4 &  2.6365897(-2) &  2.6354689(-2)    \\
           6 &  5.9305709(-2) &  5.9280563(-2)    \\
           8 &  0.1053872     &  0.1053425        \\
          10 &  0.1645720     &  0.1645023        \\
          12 &  0.2368039     &  0.2367037        \\
          14 &  0.3220005     &  0.3218648        \\
          16 &  0.4200394     &  0.4198628        \\
          18 &  0.5307267     &  0.5305039        \\
          20 &  0.6537226     &  0.6534503        \\
          22 &  0.7883219     &  0.7879974        \\
          24 &  0.9322464     &  0.9318770        \\
\hline
\end{tabular}
\caption{\label{fwell} First 24 odd and even quantum levels of the finite depth square well potential given by the CFM along with exact results. The numbers in parenthesis represent a power of 10. The barrier height $V_0=1$ and all eigenvalues are in Atomic units (see Appendix).} 
\end{center}
\end{table}

\begin{figure}[!htbp]
\begin{center}
\scalebox{0.3}{\includegraphics[angle=-90]{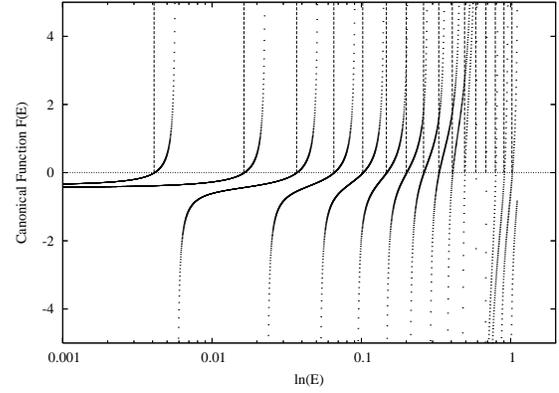}}
\end{center}
\caption{\label{tanfw} Eigenvalue function versus energy displaying the first 15 eigenvalues of 
the finite square well. The vertical lines indicate
the eigenvalue position. The eigenvalue function has an approximate
$\tan(E)$ shape versus the energy $E$.}
\end{figure}

The exact eigenvalues $E_n=\frac{\hbar^2 k_n^2}{2m}$, drawn from Landau-Lifshitz book~\cite{Landau} are given by the solutions $k_n \hspace{2mm} (n=1,2...)$ of the transcendental equation:

\begin{equation}
\sin^{-1} \frac{\hbar k_n}{\sqrt{2mV_0}}= \frac{1}{2} (n \pi -k_na); \hspace{2mm}  0 \le k_n \le \sqrt{\frac{2m V_0}{\hbar^2}}
\end{equation}

The number of levels gives us $a$ the well width in the following way: since the 
$\sin^{-1}$ term is bounded by $\frac{\pi}{2}$, the largest level index $n_{max}$ is given by 
$n_{max}=1+\frac{a}{\pi} \sqrt{\frac{2m V_0}{\hbar^2}}$, hence $a=\frac{\pi \hbar (n_{max}-1)}{\sqrt{2m V_0}}$.

\subsection{The harmonic oscillator}
The harmonic oscillator potential defined by: $V(x)=\frac{1}{2}k x^2 $ is symmetric with respect to the origin
$x=0$ implying as before that the solutions are given by odd and even parity modes. The boundaries
for this problem are: $x_1=-\infty, x_2=0$. 

In the odd-mode case (null wavefunctions at both boundaries $x_1=-\infty,x_2=0$):

\begin{eqnarray}
l_{-}(E)= \lim_{x \rightarrow -\infty} -\frac{\alpha(E;x)}{\beta(E;x)}; \nonumber \\
l_{+}(E)= \lim_{x \rightarrow 0} -\frac{\alpha(E;x)}{\beta(E;x)}; \nonumber \\
F_o(E)=l_{+}(E)-l_{-}(E)
\end{eqnarray}

whereas in the even-mode case we have (null wavefunction at left boundary $x_1=-\infty$, null 
derivative at right boundary $x_2=0$): 
\begin{eqnarray}
l_{-}(E)= \lim_{x \rightarrow -\infty} -\frac{\alpha(E;x)}{\beta(E;x)}; \nonumber \\
l_{+}(E)= \lim_{x \rightarrow 0} -\frac{\alpha'(E;x)}{\beta'(E;x)}; \nonumber \\
F_e(E)=l_{+}(E)-l_{-}(E) 
\end{eqnarray}

Solving successively $F_o(E)$ and $F_e(E)$ for the odd modes and the even modes, we get the following table~\ref{harmonic}.

\begin{table}[!htbp]
\begin{center}
\begin{tabular}{ c c c }
\hline  Index &  CFM & Exact  \\ 
\hline
           1 &  5.8842082(-2) &  5.8823533(-2)    \\
           3 &  0.1372950     &  0.1372549        \\
           5 &  0.2157480     &  0.2156863        \\
           7 &  0.2941877     &  0.2941177        \\
           9 &  0.3726121     &  0.3725490        \\
          11 &  0.4510336     &  0.4509804        \\
          13 &  0.5294515     &  0.5294118        \\
          15 &  0.6078746     &  0.6078432        \\
          17 &  0.6863770     &  0.6862745        \\
          19 &  0.7654451     &  0.7647059        \\
          21 &  0.8468009     &  0.8431373        \\
          23 &  0.9335056     &  0.9215686        \\
          25 &   1.028182     &   1.000000        \\
\hline
           0 &  1.9617772(-2) &  1.9607844(-2)    \\
           2 &  9.8067097(-2) &  9.8039217(-2)    \\
           4 &  0.1765225     &  0.1764706        \\
           6 &  0.2549707     &  0.2549020        \\
           8 &  0.3334008     &  0.3333333        \\
          10 &  0.4118234     &  0.4117647        \\
          12 &  0.4902429     &  0.4901961        \\
          14 &  0.5686612     &  0.5686275        \\
          16 &  0.6471026     &  0.6470588        \\
          18 &  0.7257724     &  0.7254902        \\
          20 &  0.8056628     &  0.8039216        \\
          22 &  0.8892854     &  0.8823529        \\
          24 &  0.9797482     &  0.9607844        \\
\hline
\end{tabular}
\caption{\label{harmonic} Ground state (zero index) and first twenty-five odd and even excited states of the harmonic oscillator potential given by the CFM along with exact results. The numbers in parenthesis represent a power of 10. The oscillator elastic constant was 
chosen such that level 25 had value 1 in Atomic units. All eigenvalues are in Atomic units (see Appendix).} 
\end{center}
\end{table}

\begin{figure}[!htbp]
\begin{center}
\scalebox{0.3}{\includegraphics[angle=-90]{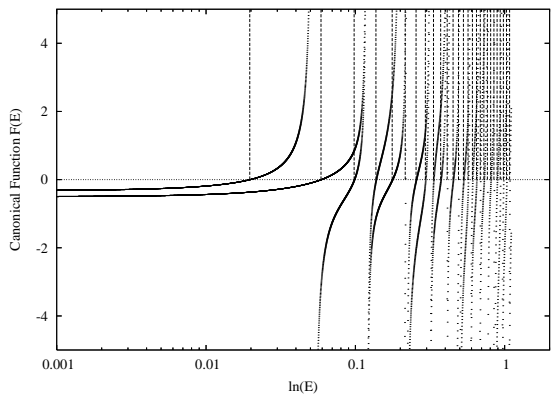}}
\end{center}
\caption{\label{tanho} Eigenvalue function versus energy displaying the first 26 eigenvalues of 
the harmonic oscillator potential. The vertical lines indicate
the eigenvalue position. The eigenvalue function has an approximate
$\tan(E)$ shape versus the energy $E$.}
\end{figure}

The exact eigenvalues $E_n=\hbar \omega_0 (n+\frac{1}{2}) $~\cite{Landau} allow us to pick the energy 
of the highest level as 1 (in Atomic units) from which we select the value of $\omega_0= \sqrt{\frac{k}{m}}$
and hence the elastic constant $k$.

\subsection{Periodic potential: The Kronig-Penney problem}
\begin{figure}[!htbp]
\begin{center}
\scalebox{0.3}{\includegraphics[angle=0]{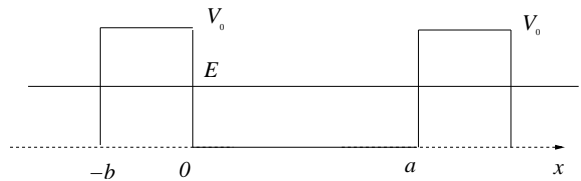}}
\end{center}
\caption{\label{Kronig} Periodic piecewise constant potential $V(x)$ displaying alternating regions of $V=0$ and $V=V_0$ 
with periodicity $a+b$. The energy bands are obtained for $E < V_0$. In the case we
let $V_0 \rightarrow \infty$ and $ b\rightarrow 0$ the barriers become delta functions sitting on a periodic lattice
with parameter $a$.}
\end{figure}

The Kronig-Penney potential is often used in the description of the electronic properties of crystals.
It is based on a piecewise constant potential (see fig.~\ref{Kronig}) for which 
we can apply the same methodology of marching to
the left and to the right and comparing corresponding ratios in order to get the eigenvalues. 
The latter are now dispersive which means they depend on a wavevector reflecting the
translational symmetry of the system (Bloch theorem). 
The CFM must be extended to the complex case since previously all the wavefunctions we use
and derive were real. It is straightforward to extend the marching method as well to the complex
wavefunction case as explained below.

Defining a unitcell with extreme boundaries $-b$ and $a$ we write the general CFM definitions
in the complex case:
\begin{eqnarray}
y(-b)&=& y(x_0) \alpha(E;-b)  + y'(x_0)\beta(E;-b)  \nonumber \\
y'(-b)&=& y(x_0) \alpha'(E;-b)  + y'(x_0)\beta'(E;-b)\nonumber \\
y(a)&=& y(x_0) \alpha(E;a)  + y'(x_0)\beta(E;a)  \nonumber \\
y'(a)&=& y(x_0) \alpha'(E;a)  + y'(x_0)\beta'(E;a)
\label{periodic}
\end{eqnarray}

The energy $E$ is considered as smaller than $V_0$.
Using Bloch theorem~\cite{Kittel}, in the above equations:
\begin{equation}
y(a)= y(-b) \exp[ik(a+b)], y'(a)= y'(-b) \exp[ik(a+b)],
\end{equation}
we get the complex ratio functions:
\begin{eqnarray}
l_{-}(E)= \left[ \frac{y'(x_0)}{y(x_0)}\right]_{-}= \frac{\gamma \alpha(E;-b)-\alpha(E;a)}{\beta(E;a)- \gamma \beta(E;-b)}; \nonumber \\
l_{+}(E)= \left[ \frac{y'(x_0)}{y(x_0)}\right]_{+}= \frac{\gamma \alpha'(E;-b)-\alpha'(E;a)}{\beta'(E;a)- \gamma \beta'(E;-b)}; \nonumber \\
F(E)=l_{+}(E)-l_{-}(E)
\end{eqnarray}

where $\gamma=\exp[ik(a+b)]$. This yields the dispersion relation for the energy 
eigenvalue $\epsilon_n(k)$ with $n$ the band index:
\begin{eqnarray}
[\gamma \alpha(E;-b)-\alpha(E;a)][\beta'(E;a)- \gamma \beta'(E;-b)] -   \nonumber \\
 \hspace{5mm}  [\beta(E;a)- \gamma \beta(E;-b)] [\gamma \alpha'(E;-b)-\alpha'(E;a)]=0
\end{eqnarray}

We compare the above to the standard dispersion relation~\cite{Kittel}:
\begin{equation}
\frac{Q^2-\kappa^2}{2Q\kappa}\sinh(Qb)\sin(\kappa a)+ \cosh(Qb)\cos \kappa a = \cos k(a+b)
\end{equation}

where $Q$ is defined as the (pure imaginary) wavevector inside the potential
barrier. Recall that the energy $E < V_0$ hence the wavefunction within the
barrier is of the form $\exp(\pm Qx)$, i.e. when $V(x)=V_0$,
 $V_0=~\frac{\hbar^2 Q^2}{2m_e}~+~\frac{\hbar^2 \kappa^2}{2m_e}$ 
whereas $\kappa$ is the 
real free wavevector outside the barrier i.e. when $V(x)=0$ the 
wavefunction is of the form $\exp(\pm i \kappa x)$).

When we let $V_0 \rightarrow \infty$ and $ b\rightarrow 0$ 
such that $V_0 b$ remains finite, the piecewise constant potential $V(x)$ is transformed 
into a periodic array of $\delta$ functions $g \delta(x-na)$ with lattice parameter $a$.
$g$ is the strength of the delta function potential and $n \in Z$ a relative integer. 

We can formally write the potential as:
\begin{equation}
V(x)= \sum_{n=-\infty}^{n=+\infty} g \delta(x-na) 
\end{equation}
and consider a single interval extending over the unit cell with boundaries $x_1=0$ 
and $x_2=a$. 
Since at the left boundary $x_1=0$ we have a $\delta$ function potential, 
standard quantum mechanics~\cite{Merzbacher} tell us that the wavefunction 
derivative $y'(x)$ jumps across the $\delta$ function potential, such that:
\begin{equation}
y'(0^+) -y'(0^-) = g\frac{2m_e}{\hbar^2} y(0)
\end{equation}
Bloch theorem~\cite{Kittel} transforms this equation into:
\begin{equation}
y'(0^+) -y'(a^-)\exp(-ika) = g\frac{2m_e}{\hbar^2} y(0)
\label{jump}
\end{equation}
The left ratio (complex) is thus obtained as:
\begin{eqnarray}
\hspace{-2cm}  l_{-}(E)= \frac{y'(x_0)}{y(x_0)} =  \nonumber \\
\frac{-\alpha'(E;0) + \alpha'(E;a)\exp(-ika) + g\frac{2m_e}{\hbar^2} 
\alpha(E;0)}{\beta'(E;0) - \beta'(E;a)\exp(-ika) - g\frac{2m_e}{\hbar^2} \beta(E;0)}
\label{krminus}
\end{eqnarray}

The right ratio is determined from Bloch theorem linking the right boundary $x_2=a$ to the
left boundary $x_1=0$: 
$y(a)=y(0) \exp(ika)$:
\begin{equation}
l_{+}(E)= \frac{y'(x_0)}{y(x_0)} = \frac{\alpha(E;a) -\exp(ika) \alpha(E;0)}{\exp(ika) \beta(E;0) -\beta(E;a)} 
\label{krplus}
\end{equation}

The dispersion relation is obtained as before from the zeroes of: 
\begin{eqnarray}
F(E)=l_{+}(E)-l_{-}(E)=\frac{\alpha(E;a) -\exp(ika) \alpha(E;0)}{\exp(ika) \beta(E;0) -\beta(E;a)} \nonumber \\
 \hspace{-5mm} - \frac{g\frac{2m_e}{\hbar^2} -\alpha'(E;0) + \alpha'(E;a)\exp(-ika) \alpha(E;0)}
{\beta'(E;0) - \beta'(E;a)\exp(-ika) - g\frac{2m_e}{\hbar^2} \beta(E;0)}
\label{dispCFM}
\end{eqnarray}

Indeed, the dispersion relation~\cite{Kittel} obtained from the limiting process 
$V_0 \rightarrow \infty$ and $ b\rightarrow 0$ is~\cite{Kittel}:
\begin{equation}
\frac{Q^2 b}{2\kappa}\sin(\kappa a)+ \cos \kappa a = \cos ka
\label{disp1}
\end{equation}

can be straightforwardly obtained from the derivative jump condition (eq.~\ref{jump}) and
Bloch theorem $y(a)=y(0) \exp(ika)$. Starting with the wave function 
 $y(x)=A \exp(i\kappa x) + B \exp(-i\kappa x)$, defined over the unit cell $x \in ]0,a[$
 and using both aforementioned conditions yields the dispersion relation:
\begin{equation}
\frac{m_e g}{\kappa \hbar^2}\sin(\kappa a)+ \cos \kappa a = \cos ka
\label{disp2}
\end{equation}

Comparing both dispersion relations yields finally the value of the strength of 
the $\delta$ function potential as $g=\frac{Q^2 b \hbar^2}{2 m_e}$.

\begin{figure}[!htbp]
  \begin{center}
    \begin{tabular}{ccc}
  \resizebox{30mm}{!}{\includegraphics[angle=0]{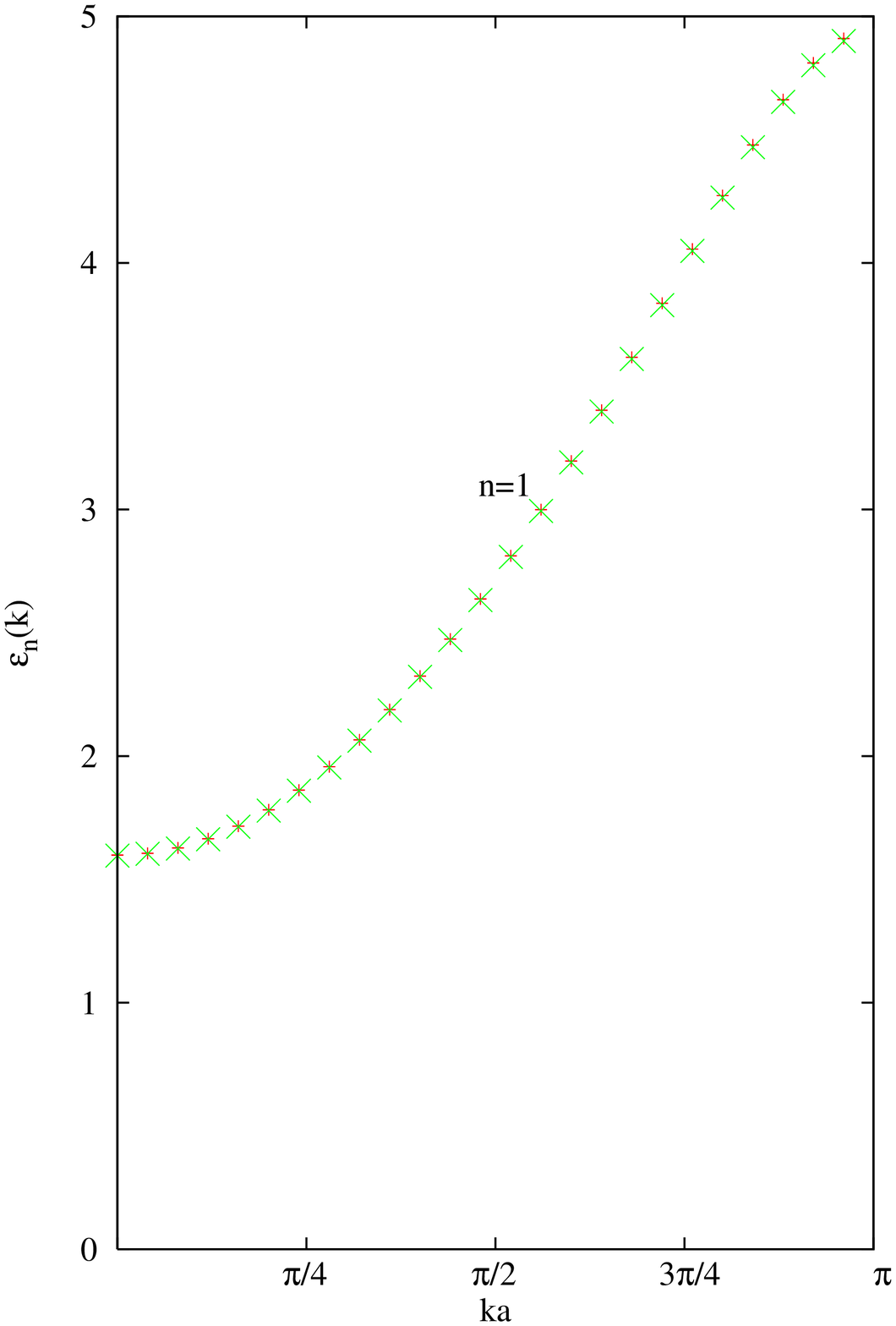}}  
                  &
   \resizebox{30mm}{!}{\includegraphics[angle=0]{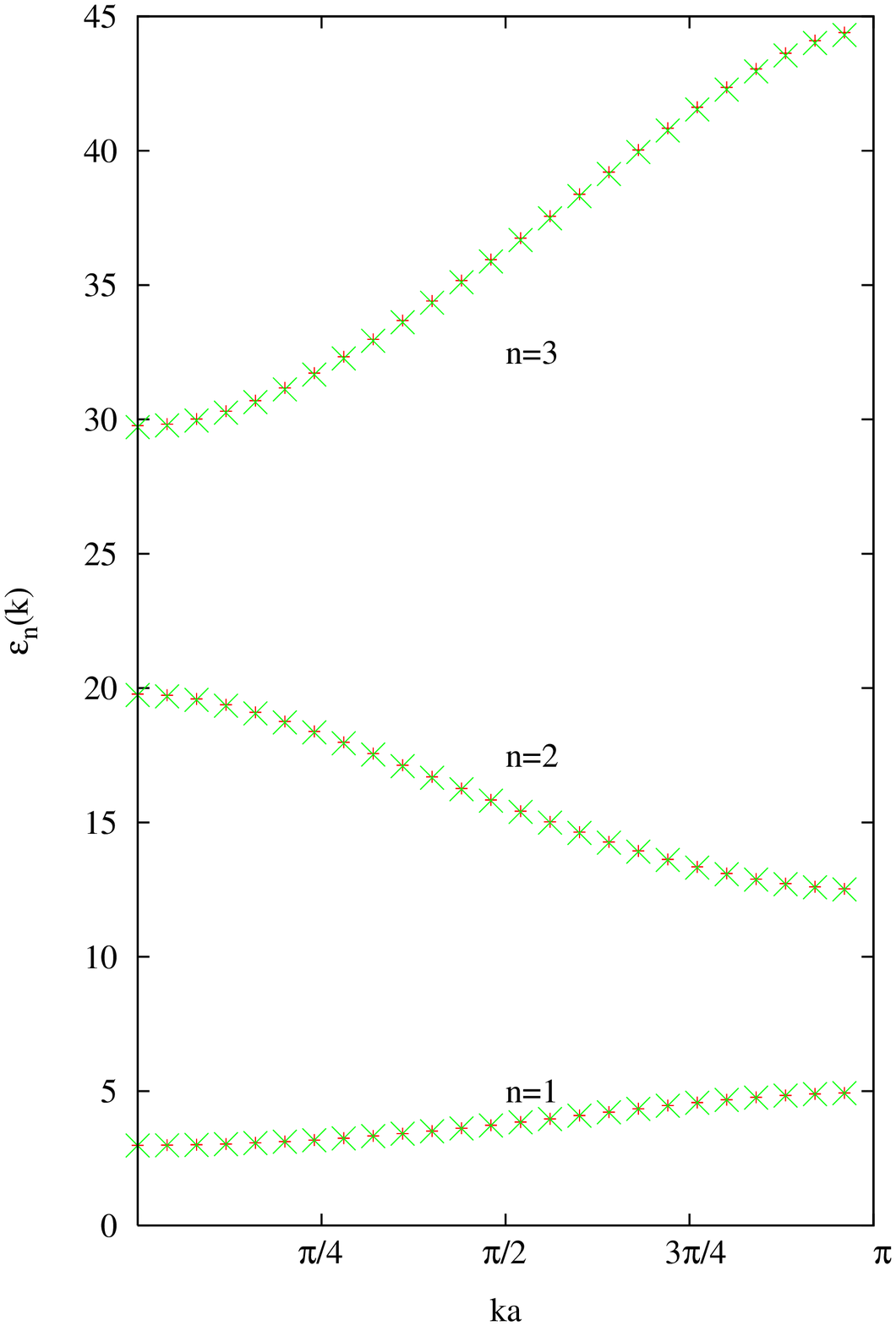}}
               &
   \resizebox{30mm}{!}{\includegraphics[angle=0]{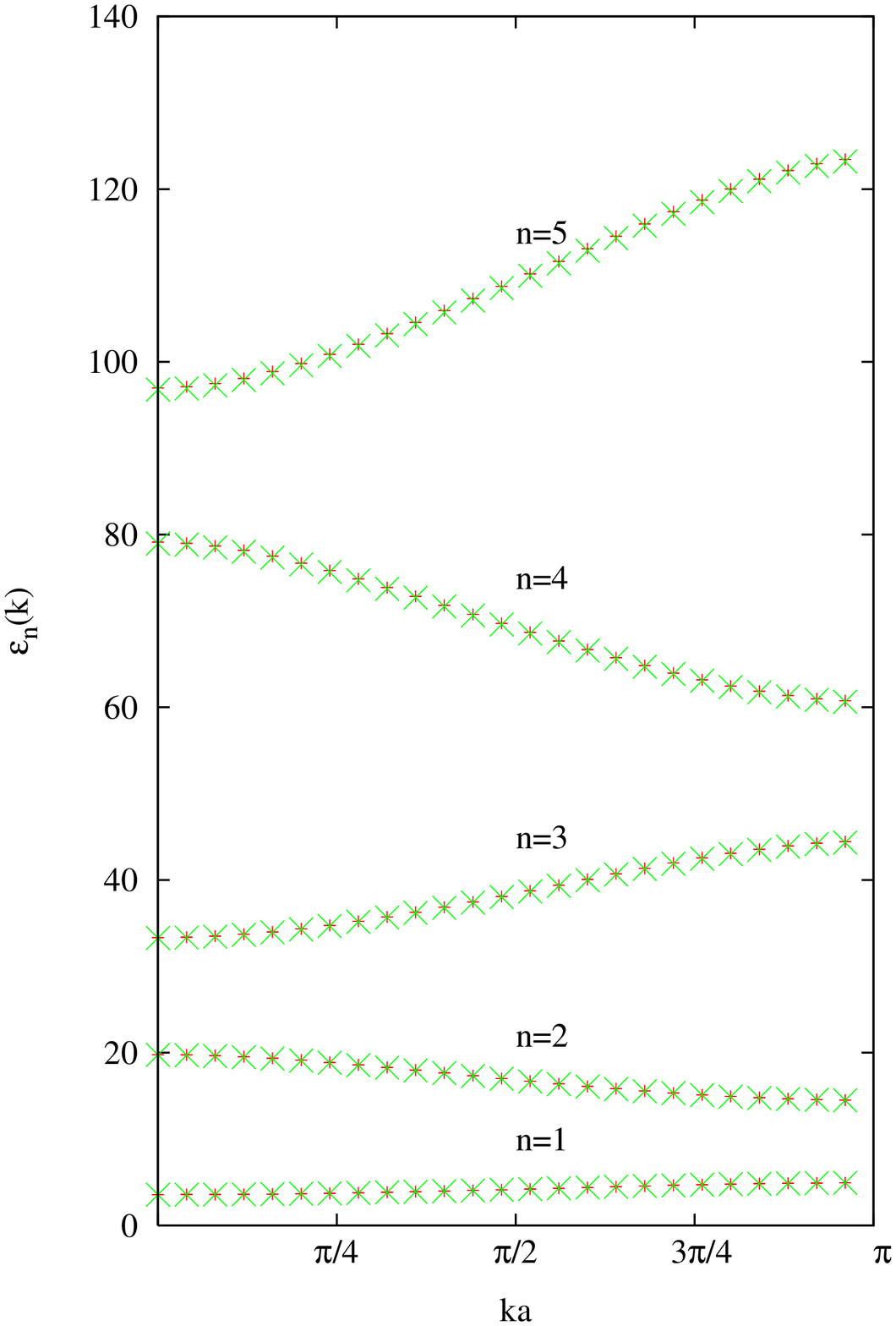}} \nonumber \\
   (a)  &  (b)  &  (c) 
    \end{tabular}
    \caption{\label{penney} (Color on line) Exact bands (in green or "$\times$") for the Kronig-Penney model and comparison with the 
CFM bands (in red or "+") obtained from the dispersion relation obtained from eq.~\ref{dispCFM} and
\ref{krminus}. (a) is for a single band, (b) and (c) are for 3 and 5 bands respectively
with $a=2.22, 6.66, 11.12$ Atomic units.
In all cases, the value of $x_0$ is 1 and the strength of the potential $g=1$ in Atomic units.}
\end{center}
\end{figure}

In figure~\ref{penney}, exact bands are compared to the CFM bands. The lattice parameter 
in each case is determined from the number of bands $n_{B}$ we want to calculate according to
the formula: $a=\frac{n_{B}\pi \hbar}{\sqrt{2 m_e}} $ since the largest wavenumber is
$k_{max}=\frac{n_{B}\pi}{a}$ and we select the largest energy 
$E_{max}=\frac{\hbar^2 k^2}{2m_e}$ as 1. This is why we use  
$a=2.22, 6.66, 11.12$ Atomic units for the $n_B=1,3,5$ respective number of bands.
It is remarkable to observe how the CFM results lie exactly on top of the exact results.
The starting value $x_0$ is chosen in a way such that we get the right number of bands 
$n_{band}$.
Spurious bands might appear due to a bad starting value $x_0$ because the
nature of the CFM dispersion relation~\ref{dispCFM} differs with respect to the 
dispersion relations eq.~\ref{disp1} and eq.\ref{disp2}. 
The latter eq.\ref{disp2} allow the exact determination
of the free wavevector $\kappa$ from a given Bloch wavevector $k$ and the exact band energy is
obtained from $E=\frac{\hbar^2 \kappa^2}{2m_e}$. In sharp contrast, the CFM dispersion
relation~\ref{dispCFM} yields directly the band energy without going through
the determination of an intermediate wavevector $\kappa$.

\subsection{Double well potential over an infinite interval}
Double-minimum Potential Well (DPW) problems defined over the 
semi-infinite interval $[0,\infty[$ are interesting to solve as they arise
in many  areas of Atomic, Molecular and even Solid State physics.
When two-dimensional electron layers (such as in heterostructures involving semiconductors) 
are placed in perpendicular electric and magnetic fields, a potential well with
two minima, for electronic motion normal to the surface, arises. \\

A DPW can be symmetric or asymmetric and one has to adapt in each case
the appropriate BC imposed by the CFM.

A symmetric DPW is the Double Gaussian potential
investigated by Hamilton and Light \cite{Ham} given by:
$$
    V(x)=-D [ \exp(-\Omega {(x-r_{a})}^2)+\exp(-\Omega {(x+r_{a})}^2)]
$$

The values of the parameters: $D, \Omega, r_{a}$ are respectively: 12.0,0.1,5.0 in standard atomic
units (see Appendix) such that $\hbar=1, m_e=1$.

An elaborate method used by Hamilton and Light \cite{Ham} based on Distributed
Gaussian Basis sets borrowed from Quantum Chemistry Techniques gives the eigenvalues
listed in table~\ref{dgw}. The CFM results for all the 24 levels in table~\ref{dgw} proves
once again  that it is able to find all levels with speed and accuracy from a simple
marching approach.  

\begin{table}[!hb]
\begin{center}
\begin{tabular}{ c c l }
\hline
  Index & CFM &  Hamilton and Light \\
\hline
1  &      -11.250 421 409     & -11.245 199 313 \\
 3    &   -9.779 225 834     &  -9.773 496 902  \\
  5   &  -8.387 719 137      &  -8.381 307 491  \\
   7  &  -7.079 412 929       & -7.072 038 846  \\
 9    &  -5.858 811 221      & -5.849 940 0 \\
  11  &   -4.732 171 001    & -4.720 509 6  \\
    13&   -3.709 113 861   & -3.690 475 6  \\
  15  &   -2.801 628 760    & -2.763 219 7  \\
    17&   -2.000 566 637    & -1.924 577 \\
   19 &   -1.255 332 005    & -1.149 254 \\
 21   &   -0.561 216 170    & -0.457 88 \\
   23 &   -0.045 810 537   & -0.003 41 \\   
\hline
  0 &  -11.250 418 469      &  -11.245 199 313   \\	
  2  & -9.779 202 594       &  -9.773 496 902	 \\	
  4   & -8.387 701 732       & -8.381 307 510 \\		
   6 & -7.079 415 041     & -7.072 039 562 \\		
  8   & -5.858 805 474      & -5.849 958 02  \\	
  10   & -4.732 231 858     & -4.720 829 36 \\	
 12  &  -3.709 907 559      & -3.694 518 38 \\
  14  &  -2.807 436 691     & -2.798 251 92  \\	
 16    & -2.022 064 904     & -2.089 661 3 \\		
 18   &  -1.293 090 067      & -1.462 202 9 \\	
  20  &  -0.601 483 056     & -0.771 081  \\	
  22  &  -0.067 153 689      & -0.177 181 \\
      \hline
\end{tabular}
\end{center}
\caption{\label{dgw} Computed odd and even eigenvalues for the symmetric double Gaussian well potential.
The numbers at left are the levels computed 
with the CFM; on the right the levels obtained by Hamilton and Light \cite{Ham}. Note
the deterioration of accuracy of the Hamilton and Light results as the index increases 
because of the approach of the continuum.}
\end{table}

The asymmetric DWP introduced by Johnson consists of the sum of 
a Morse (see next section) and a Gaussian potentials such that:

$$
V(x)=D{[1-\exp(-B(x-r_{a}))]}^2+A \exp{(-C(x-r_{b})}^2) 
$$       

The values of the parameters $A,B,C,D,r_a,r_b$ are (following Johnson~\cite{Johnson})
 in (cm$^{-1}$, \AA \hspace{1mm} system of units) are: 10$^4$ cm$^{-1}$,    1.54 \AA$^{-1}$, 
 200.0 \AA$^{-2}$,    31250.0 cm$^{-1}$,   1.5  \AA,    1.6  \AA  \hspace{1mm} respectively.

Eigenvalues for the asymmetric double minimum 
potential problem are given in table~\ref{DWP2} and  a
comparison between Johnson's~\cite{Johnson} results and the CFM are displayed below.

\begin{table}[!htbp]
\begin{center}
\begin{tabular}{c c c}
\hline Index& Johnson & CFM \\
\hline
0 & 1302.500  &         1302.498 972    \\
1 & 3205.307  &  	3205.303 782   \\ 
2 & 4227.339  &  	4227.336 543    \\
3 & 5144.251  &  	5144.243 754  \\
4 & 6064.241  &  	6064.225 881   \\ 
5 & 7092.679  &  	7092.664 815  \\
6 & 7614.622  &  	7614.603 506  \\
7 & 8911.545  &  	8911.513 342    \\ 
8 & 9095.696  &  	9095.679 497   \\  
9 & 10208.350  &    10208.318 142    \\ 
10 & 10869.289  &    10869.255 077    \\ 
11 & 11482.475  &    11482.457 956    \\ 
12 & 12353.799  &    12353.766 422    \\  
13 & 12972.473  &    12972.453 117    \\ 
14 & 13690.455  &    13690.436 602   \\  
15 & 14435.350  &    14435.321 044   \\ 
\hline
\end{tabular}
\caption{\label{DWP2} Eigenvalues in cm$^{-1}$ of the Johnson asymmetric DWP consisting of the
sum of a Morse and a Gaussian potentials 
$V(r)=D{[1-\exp(-B(x-r_{a}))]}^2+A \exp{(-C(x-r_{b})}^2)$.
with $A$= 10$^4$ cm$^{-1}$,   $B$= 1.54 \AA$^{-1}$, 
$C$= 200.0 \AA$^{-2}$,   $D$= 31250.0 cm$^{-1}$, 
and $r_a$ =  1.5  \AA,  $r_b$=  1.6  \AA .
Johnson~\cite{Johnson} results are compared to the CFM.}
\end{center}
\end{table}

\section{The Canonical Function Method and the 3D Radial Schr\"{o}dinger Equation}

After having discussed the CFM in the 1D case, we move on to the treatment of the Radial Schr\"{o}dinger
Equation (RSE). The mathematical difficulty of the RSE lies in the fact it is a singular boundary value problem (SBVP). 
problem. This stems from the boundary conditions over the infinite interval $[0, \infty[$ , with the 
double requirement of regularity near the origin ($r \sim 0$) where the potential is large and 
near infinity ($r \rightarrow \infty$) where the potential is very small.
The CFM turns it into a regular initial value problem and allows the full determination of the 
spectrum of the Schr\"{o}dinger operator bypassing the evaluation of the eigenfunctions.\\

The partial wave form of the RSE is written as:

\begin{eqnarray}
-\frac{\hbar^2}{2\mu}{\frac{d^2 R_{l} ( E; r )}{d r^2}}  + 
\left[{V( r ) +  \frac{\hbar^2}{2\mu}\frac{ l (l + 1 )}{r^2}}\right] R_{l} (E; r ) = \nonumber \\
\hspace{3cm} E R_{l} (E; r )
\end{eqnarray}

where $\mu$ is the reduced mass and $R_{l} (E; r )$ is the reduced probability amplitude for 
orbital angular momentum $l$ and eigenvalue $E$.

The BC are:
\begin{equation}
\lim_{r \rightarrow 0} R_{l} (r) =0; \lim_{r \rightarrow +\infty} R_{l} (r) =0
\end{equation}

The CFM  consists of writing the general  solution $y(r)$ representing the probability
amplitude $R_{l} (E; r )$ as a function of the radial distance 
$r$ in terms of two linearly independent basis functions $\alpha(E;r)$ and $\beta(E;r)$ for some energy $E$. \\
Generally, the RSE is rewritten in a system of units such that $\hbar=1, 2 \mu=1$ 
(see Appendix on units):
\begin{equation}
\frac{d^2 y(r)}{d r^2} = \left[{V( r ) +  \frac{ l (l + 1 )}{r^2}-E }\right] y(r) 
\label{schro}
\end{equation}
At a selected distance $r_0$, a well defined set of initial conditions are satisfied by the 
canonical functions and their derivatives ie: $\alpha(E;r_0)=1$ with $\alpha'(E;r_0)=0$ 
and $\beta(E;r_0)=0$ with $\beta'(E;r_0)=1$. Thus we write as in the 1D case:

\begin{equation}
y(r)= y(r_0) \alpha(E;r)  + y'(r_0)\beta(E;r)
\end{equation}

The method of solving the RSE is to proceed from $r_0$ simultaneously towards the origin
($r \rightarrow 0$) and towards infinity ($r \rightarrow \infty$). 

When the integration is performed, the ratio of the $r$ dependent canonical functions is monitored until
saturation with respect to $r$ is reached at both limits ($r \rightarrow 0$ and $r \rightarrow \infty$).
The saturation of the $\frac{\alpha(E;r)}{\beta(E;r)}$ ratio with $r$ yields a position independent 
eigenvalue function $F(E)$:

\begin{equation}
F(E)=l_{+}(E)-l_{-}(E)= \left[ \frac{y'(r_0)}{y(r_0)}\right]_{+} - \left[\frac{y'(r_0)}{y(r_0)}\right]_{-}
\end{equation}

The $\tan(E)$ shape of $F(E)$ provides a deep insight into the physical significance
of the CFM method. The latter transforms a SBVP from the open interval $[0, \infty[$ 
to a finite interval $[r_{left}, r_{right}]$ defined by the saturation coordinates of the
ratio functions. This means the CFM maps an arbitrary potential $V(r)$ onto the
infinite square well problem in the finite interval $[r_{left}, r_{right}]$  resulting
in an eigenvalue function  $F(E)$ with a $\tan(E)$ pattern as we saw in Section II 
(see also ref.~\cite{Johnson}). 

\subsection{The Hydrogen atom spectrum}

The Coulomb potential is a crucial case to test the accuracy and reliability 
of the CFM given by the Hydrogen atom problem.

The CFM results are shown in Table.~\ref{coul} along with the exact analytical
results and it is remarkable to notice that all digits (calculated by CFM and analytically) are
rigorously same.\\

\begin{table}[!htbp]
\begin{center}
\begin{tabular}{ c c c }
\hline
Index& CFM (Ry) & Exact (Ry) \\
\hline
 1   &-1.00000   &   -1.00000 \\
   2  &-0.250000    &  -0.250000 \\
   3  &-0.111111    & -0.111111 \\
   4 &  -6.25000(-2)    &  -6.25000(-2) \\
   5  & -4.00000(-2)    &   -4.00000(-2) \\
   6  & -2.77778(-2)    &   -2.77778(-2) \\
   7  & -2.04082(-2)    &   -2.04082(-2) \\
   8 &  -1.56250(-2)    &   -1.56250(-2) \\
   9  & -1.23457(-2)    &   -1.23457(-2) \\
   10 &  -1.00000(-2)    &   -1.00000(-2) \\
   11 &  -8.26446(-3)    &   -8.26446(-3) \\
   12  & -6.94444(-3)    &   -6.94444(-3) \\
   13  & -5.91716(-3)    &   -5.91716(-3) \\
   14 &  -5.10204(-3)    &   -5.10204(-3) \\
   15 &  -4.44445(-3)    &   -4.44445(-3) \\
   16 &  -3.90625(-3)    &   -3.90625(-3) \\
   17 &  -3.46021(-3)    &   -3.46021(-3) \\
   18 &  -3.08642(-3)    &   -3.08642(-3) \\
   19 &  -2.77008(-3)    &   -2.77008(-3) \\
   20 &  -2.50000(-3)    &   -2.50000(-3) \\
   21 &  -2.26757(-3)    &   -2.26757(-3) \\
   22 &  -2.06612(-3)    &   -2.06612(-3) \\
   23 &  -1.89036(-3)    &   -1.89036(-3) \\
   24 &  -1.73611(-3)    &   -1.73611(-3) \\
\hline
\end{tabular}
\end{center}
\caption{\label{coul} Energy levels of the Hydrogen atom. Middle column values are the CFM results whereas
the last column values are the corresponding exact analytically obtained values. 
The numbers in parenthesis represent a power of 10.}
\end{table}

\subsection{The Morse potential}

The classical Morse potential is the simplest model for the evaluation of cell
vibrational spectra of  diatomic molecules. 

The Morse potential is given by:

\begin{equation}
        V(r)=D {[ 1  -\exp(-a \{r-r_e \}) ]}^2 -D
\end{equation}

with the values $D,a,r_e$ equal respectively to  188.4355,  0.711248,   1.9975 in atomic units (see Appendix).
The analytic expression for the levels is:

\begin{equation}
E_{n}=- \frac{a^2 \hbar^2}{2 \mu}(\frac{\sqrt{2\mu D}}{a} -n- \frac{1}{2})^2,
\end{equation}

with max $n \le \frac{\sqrt{2\mu D}}{a} - \frac{1}{2}$. Hence the number of levels
 is given by: $N=\frac{\sqrt{2\mu D}}{a}- \frac{1}{2}$.

\begin{figure}[!htbp]
\begin{center}
\scalebox{0.3}{\includegraphics[angle=-90]{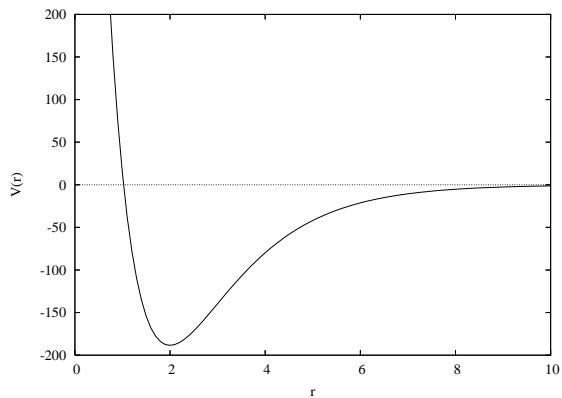}}
\end{center}
\caption{\label{morsepot} Morse potential  $V(r)=D {[ 1  -\exp(-a \{r-r_e \}) ]}^2 -D$ with parameters 
$D= 188.4355, a= 0.711248,  r_e= 1.9975$.} 
\end{figure}

\begin{figure}[!htbp]
\begin{center}
\scalebox{0.3}{\includegraphics*[angle=-90]{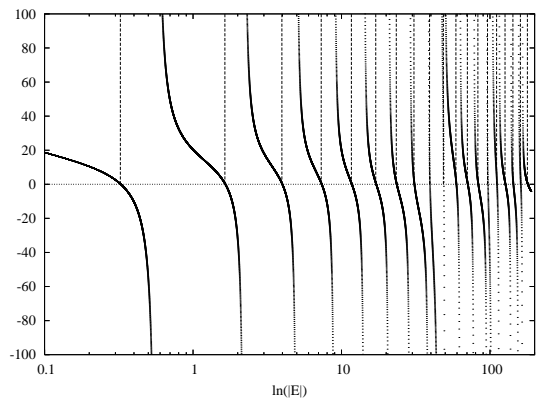}}
\end{center}
\caption{\label{morse} Behavior of the eigenvalue function $F(E)$ with absolute value of energy on a
semi-log scale for the Morse potential.} 
\end{figure}

Working with units such that $\hbar=1$ and $2\mu=1$,
the Morse potential and the eigenvalue function $F(E)$ are displayed in fig.\ref{morsepot} 
and fig.~\ref{morse} respectively.
Table~\ref{morset} contains the levels calculated by CFM and compared to the analytical 
analytical results.
As in all previous cases, the agreement is perfect
and the full set of levels ($N=19$) are found as predicted analytically.

\begin{table}[!htbp]
\begin{center}
\begin{tabular}{c c c}
\hline
Index & CFM  & Exact \\
\hline
 1 & -178.798248 &  -178.798538 \\
 2 & -160.282181 &  -160.283432 \\
 3 & -142.778412 &  -142.78006 \\
 4 &  -126.287987 &  -126.288445 \\
 5 &  -110.807388 &  -110.808578 \\
 6 &  -96.3395233  & -96.3404541 \\
 7 &  -82.8832169 &  -82.884079 \\
 8  & -70.4389801 &  -70.4394531 \\
 9 & -59.0056    &  -59.0065727 \\
 10 & -48.5851288 &  -48.5854378 \\
 11 &  -39.1754532 &  -39.1760521 \\
 12 & -30.77771  &   -30.7784157 \\
 13 & -23.3919983 &   -23.3925247 \\
 14 & -17.0183048  &  -17.018383 \\
 15 & -11.6557436 &   -11.6559868 \\
 16 &  -7.3050122 &  -7.30533791 \\
 17 & -3.9661877 &   -3.9664371 \\
 18 & -1.6390723  &   -1.63928342 \\
 19 & -0.3238727 &    -0.32387724 \\
 \hline
\end{tabular}
\end{center}
\caption{\label{morset} Energy levels of the Morse potential $V(r)=D {[ 1  -\exp(-a \{r-r_e \}) ]}^2 -D$ with parameters 
$D= 188.4355, a= 0.711248,  r_e= 1.9975$. Middle column values are the CFM 
results whereas the last column values are the corresponding exact analytically obtained values.
Units are such that $\hbar=1$ and $2\mu=1$.}
\end{table}

\section{Conclusions}

The CFM is a very powerful, fast and accurate method that is able to evaluate 
the eigenvalue spectrum without having to determine first or simultaneously
the eigenfunctions.

The CFM has been tested succesfully in a variety of potentials~\cite{Tannous} 
and gives accurate results for bound and free states. 
The tunable accuracy of our method allows to evaluate eigenvalues
close to the ground state as well as close to highly excited states near
the continuum limit to a large number of digits without any extrapolation.\\

The CFM compares favorably with many different elaborate techniques 
based on expansion over basis functions (such as Gaussian~\cite{Johnson}, Quantum Chemistry 
inspired basis functions~\cite{Ham}) or functional expansions (Numerov~\cite{Numerov},
High-order Taylor~\cite{Raptis}...).
The CFM approach remains the same despite the wide variability of the mentioned problems.  

The CFM method used gives the right number of all the levels
and the variation of the eigenvalue function $F(E)$ definitely determines the
total number of levels. Generally in order to avoid potential singularities,
Taylor series expansion are made to a given order dictated by the required accuracy 
(as described in ref~\cite{Fornberg}). \\

Since the CFM bypasses the calculation of the
eigenfunctions, it avoids losing accuracy associated with the
numerical calculation specially with rapidly oscillating wave functions
of highly excited states. This is specially needed in the study of the 
sensitive problem of Rydberg states in Atomic physics or 
the determination of vibrational spectra of cold (weakly-bound) molecules...

{\bf Acknowledgements}: \\
We would like to acknowledge helpful correspondance with Jeff Cash (Imperial College),
Ronald Friedman (Purdue), Bengt Fornberg (Caltech) and John W. Wilkins (Ohio state).\\

\begin{center}
{\bf APPENDIX}
\end{center}

{\bf Atomic and other units} \\

In atomic and molecular physics, it is convenient to use the elementary
charge $e$, as the unit of charge, and the electron mass $m_e$
as the unit of mass (despite the fact that in some cases
the proton mass, $m_p$ , or the unified mass unit amu,
is more convenient). Electrostatic forces and energies in atoms are
proportional to $e^2/4\pi \epsilon_0$ , which has dimensions
$ML^3T^{-2}$, and another quantity that appears all over in
quantum physics is $\hbar$ which has dimensions $ML^2T^{-1}$
; so it is convenient to choose units of length and time such that
 $4\pi \epsilon_0=1$ and $\hbar=1$.

\begin{table}[!h]
\centering
\resizebox{80mm}{!}{
\begin{tabular}{l|c|c|c|c}
\hline
          & J & eV & Hz & cm$^{-1}$  \\
\hline
 J   & 1 &  6.24151.10$^{18}$ &  1.50919.10$^{33}$ & 5.03411.10$^{22}$  \\
\hline
 eV  & 1.60219.10$^{-19}$ & 1 & 2.41797.10$^{14}$ & 8.06547.10$^{3}$ \\
\hline
 Hz  & 6.62619.10$^{-34}$ & 4.13570.10$^{-15}$ & 1 & 3.33564.10$^{-11}$  \\
\hline
 cm$^{-1}$  & 1.96648.10$^{-23}$ & 1.23935.10 $^{-4}$ & 2.99792.10$^{10}$ & 1  \\
\hline
\end{tabular}
}
\caption{\label{units} Conversion table for the energy expressed in J, eV, Hz and cm$^{-1}$. In MKS the Joule
is preferred whereas physicists in general use eV or Hz. Spectroscopists and Chemists use rather the cm$^{-1}$. }
\end{table}

Using dimensional analysis, the atomic unit of length is: 

\begin{equation}
a_B=\frac{e^2}{m_e (e^2/4\pi \epsilon_0)},
\end{equation}

called the Bohr radius, or simply the bohr (0.529 \AA), 
because in the "Bohr model" the radius of the smallest orbit for an electron circling a
fixed proton is $(1+\frac{m_e}{m_p})a_B$.  In full quantum theory
the particles do not follow an orbit but possess wavefunctions 
and the expectation value of the
electron-proton distance in the Hydrogen ground state is exactly
$(1+\frac{m_e}{m_p})a_B$.

The atomic unit of energy is the Hartree (27.2 eV) given by:
\begin{equation}
E_h= \frac{e^2}{4\pi \epsilon_0}\frac{1}{a_B}={(\frac{e^2}{4\pi \epsilon_0})}^2\frac{m_e}{\hbar^2}
\end{equation}

The unit of time is $\hbar/E_h$.

The Hartree is twice the ground state energy of the Hydrogen atom
$\frac{1}{2}{(1+\frac{m_e}{m_p})}^{-1}E_h$ equal to the Rydberg (13.6 eV). 
In atomic and molecular spectroscopy, one uses rather the cm$^{-1}$ an energy
corresponding to a wavelength of 1cm or sometimes a frequency unit, the Hz.
We refer the reader to table~\ref{units} where conversion factors between the
different energies are given.

\end{document}